\documentclass[9pt]{amsart}
\usepackage{multicol}
\usepackage{fixltx2e}
\usepackage[paperwidth=210mm, paperheight=276mm, 
            columnsep=4mm, margin=14mm]{geometry}
\usepackage[pdftex]{graphicx}

\newif\ifscientificdata
\scientificdatafalse

\ifscientificdata
\else
 \graphicspath{{./figures/}}  
\fi

\usepackage[super, numbers]{natbib}

\usepackage[usenames,dvipsnames]{color}
\usepackage{epstopdf,gensymb,xfrac,fixltx2e,url,booktabs,siunitx,ctable,
            subfig,float,csvsimple,enumerate}
\usepackage[bookmarks,pdfdisplaydoctitle,colorlinks,
            pdfauthor={Jack Kelly},
            pdftitle={The UK-DALE dataset},
            urlcolor=BlueViolet, 
            citecolor=CadetBlue,
            linkcolor=black
            ]{hyperref}

\begin{document}

\title[The UK-DALE dataset] {The UK-DALE dataset, domestic appliance-level electricity\\
demand and whole-house demand from five UK homes}

\date{January 2015}
\author[Jack~Kelly \&  William~Knottenbelt]
{Jack~Kelly$^1$ \&  William~Knottenbelt \\
  \textit{Department of Computing, Imperial College London, London,
    SW7 2RH, UK}}
\footnotetext[1]{Email: \texttt{jack.kelly@imperial.ac.uk}; corresponding author}

\begin{abstract}

Many countries are rolling out smart electricity meters.  These
measure a home's total power demand.  However, research into consumer
behaviour suggests that consumers are best able to improve their
energy efficiency when provided with itemised, appliance-by-appliance
consumption information.  Energy disaggregation is a computational
technique for estimating appliance-by-appliance energy consumption
from a whole-house meter signal.

To conduct research on disaggregation algorithms, researchers require
data describing not just the aggregate demand per building but also
the `ground truth' demand of individual appliances.  In this context,
we present UK-DALE: an open-access dataset from the UK recording
Domestic Appliance-Level Electricity at a sample rate of 16~kHz for
the whole-house and at $\sfrac{1}{6}$~Hz for individual
appliances. This is the first open access UK dataset at this temporal
resolution. We recorded from five houses, one of which was recorded for
655~days, the longest duration we are aware of for any energy dataset
at this sample rate.  We also describe the low-cost, open-source,
wireless system we built for collecting our dataset.
\end{abstract}

\maketitle
\begin{multicols}{2}

\section{Background \& Summary}

Prudent management of electricity consumption is becoming increasingly
important. Yet studies on residential energy users show that the vast
majority are poor at estimating their whole-house energy consumption
or the energy consumption of individual
devices\cite{Kempton1982}. Residents often underestimate the energy
used by heating and overestimate the consumption of perceptually
salient devices like lights and
televisions\cite{Kempton1982}. Residents' failure to correctly
estimate energy consumption is likely to lead to higher total
consumption.

How significant is occupant behaviour in determining total energy
usage? Energy use can differ by two or three times among identical
houses with similar appliances occupied by people from similar
demographics\cite{Socolow1978,Winett1979,SeryakJ.2003}. These large
differences in energy consumption are attributed to differences in
consumption behaviour. If the house provided better feedback about
which devices used the most energy then users could adjust their
behaviour to make more efficient use of appliances.  `Smart
electricity meters' are one such feedback mechanism.

The UK Government requires that energy retailers install smart
electricity and gas meters by 2020.  The roll-out is already
underway\cite{DECC2013}. Similar smart meter roll-outs are planned in
many countries.

The business case for smart meters in Great Britain\cite{DECC2011b}
assumes that smart meters will drive savings of \pounds4.6~billion 
due to reduced energy consumption (across both
electricity and gas).  Smart meters only provide energy consumption
measurements for the entire house yet behavioural research suggests
that consumers are best able to manage their electricity consumption
when given \emph{appliance-by-appliance} information
\cite{Fischer2008}.  Energy disaggregation aims to estimate
appliance-by-appliance consumption from a smart meter signal and hence
may play an important role in realising the energy savings projected
by the smart meter business case.

Energy disaggregation\cite{Hart1992} is an active area of research
(see Armel \emph{et al.} 2013\cite{Armel2013} for a recent review).
Researchers require access to large datasets recorded in the field to
develop disaggregation algorithms but it is not practical for every
researcher to record their own dataset.  Hence the creation of open
access datasets is key to promote a vibrant research community.

Researchers at MIT led the way by releasing The Reference Energy
Disaggregation Data Set (REDD) in 2011\cite{J.ZicoKolter2011} and
more datasets have subsequently been released by researchers in the
USA\cite{Anderson2012,Barker2012,DataPort,BERDS},
Canada\cite{Makonin2013}, India\cite{IAWE,COMBED},
France\cite{Hebrail2012}, the UK\cite{HES}, Switzerland\cite{ECO}, 
Portugal\cite{Sustdata} and Italy and Austria\cite{GREEND}.

To test the performance of a disaggregation algorithm for a specific
country, it is important to have access to data from that country
because electricity usage varies significantly between countries; both
because different countries use different sets of appliances and also
because different cultures show different usage patterns.

At the time of writing, the only open-access dataset recorded in the
UK is the DECC/DEFRA Household Electricity Study\cite{HES} which has
a sample period of two minutes.  Yet this sample rate is 12 times slower
than UK smart meters which will sample every 10~seconds\cite{SMETS2}.
It is this smart meter data which will provide the input to
disaggregation algorithms so researchers require access to 10~second
data to design disaggregation algorithms for the UK (some other
countries will also use smart meters with similar sample periods).

We present the first open access UK dataset with a high temporal
resolution.  We recorded from five houses.  Every six~seconds we
recorded the active power drawn by individual appliances and the
whole-house apparent power demand.  Additionally, in three houses,
we sampled the whole-house voltage and current at 44.1~kHz
(down-sampled to 16~kHz for storage) and also calculated the active
power, apparent power and RMS voltage at 1~Hz.  In House~1, we
recorded for 655~days and individually recorded from almost every
single appliance in the house resulting in a recording of 54 separate
channels (although less channels were recorded towards the start of
the dataset).  We will continue to record from this house for the
foreseeable future.  We recorded from the four other houses for
several months; each of these houses recorded between 5 to 26~channels
of individual appliance data. Figure~\ref{fig:system_diagram} provides
an overview of the system design and Table~\ref{table:houses}
summarises the dataset.

This dataset may also be of use to researchers working on:

\begin{itemize}
  \item modelling the electricity grid.
  \item exploring the potential for automated demand response.
  \item appliance usage behaviour.
\end{itemize}

\section{Methods}




Desirable attributes of a dataset for disaggregation
include:

\begin{itemize}
  \item Simultaneously record the power drawn by
    most of the individual appliances in each house (this data can be
    used to validate the appliance-by-appliance estimates produced by
    a disaggregation system or to train the system).
  \item Record the whole-house active power (this will be the input to
    the disaggregation algorithm).
  \item Sample once every 10~seconds or faster.
  \item Record for as long as possible.
\end{itemize}

We first describe our approach to monitoring individual appliances
once every 6~seconds and then describe how we recorded whole-house
mains power at 44.1~kHz.

\subsection{Individual appliance monitoring}

\begin{figure*}
  \centering
  \ifscientificdata
    01\_system\_diagram.eps
  \else
    \includegraphics[width=\textwidth]{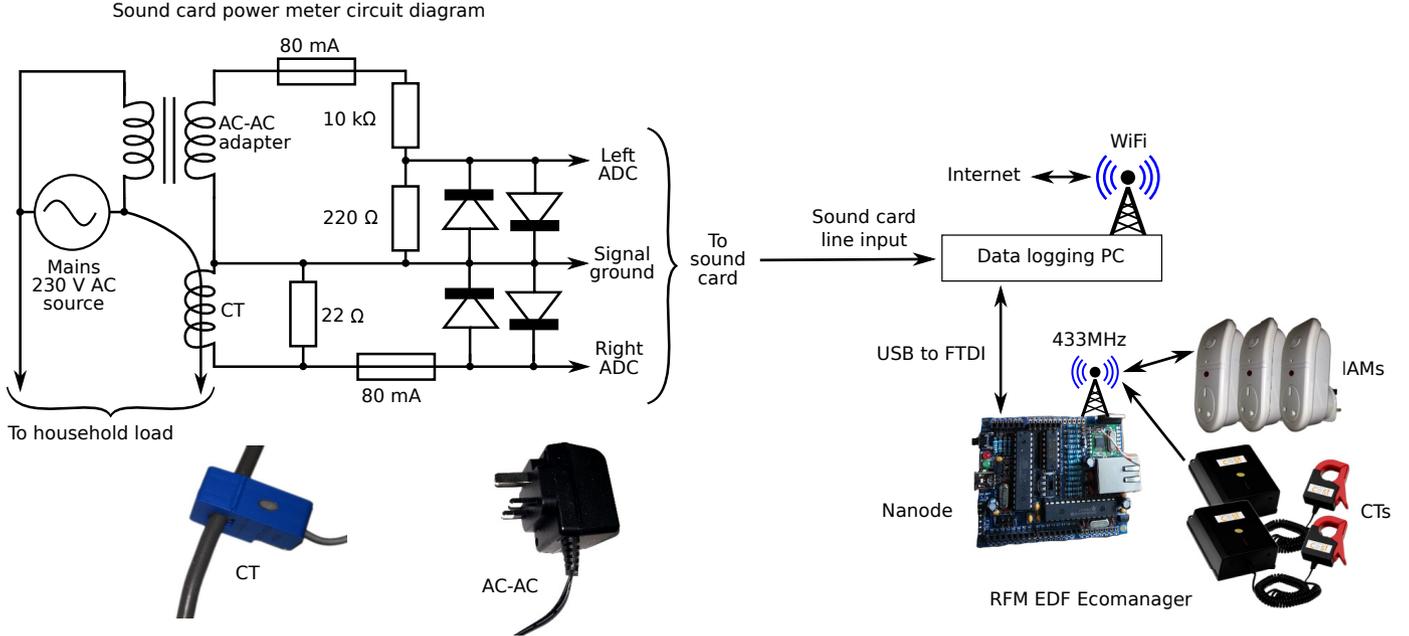}
  \fi
  \caption{\textbf{System diagram for the data collection system.}
    The system has three major components: 1) the data logging PC; 2)
    the sound card power meter and 3) the `RFM EDF Ecomanager' which
    uses a Nanode to communicate over the air with a set of individual
    appliance monitors (IAMs) and current transformer (CT) sensors.
    On the left is the circuit diagram for interfacing a sound card to
    a CT clamp and \mbox{AC-AC} adaptor to measure mains current and
    voltage, respectively.  The circuit was adapted from Robert Wall's
    work\cite{Wall}.
    Each diode is a 1N5282 (1.3~V forward voltage bias).}
  \label{fig:system_diagram} 
\end{figure*}

In UK houses such as those in our dataset, mains `rings' extend from
the fuse box.  Many sockets may share the same ring.  Hence, in order to
measure individual appliances in the UK, we must install plug-in
individual appliance monitors (IAMs) between each appliance and its
wall socket.

We used EcoManager Transmitter Plugs\cite{EcoManagerTRX} developed
by Current Cost and distributed by EDF Energy.  The standard base
station for these IAMs is the EcoManager\cite{EcoManager}.  The
EcoManager can only handle a maximum of 14 transmitter plugs and only
provides data once per minute via its serial port.  We needed up to 54
appliances monitored per house and data every 10~seconds or faster.
To achieve this, we set about building our own base station.  With the
help of others in the community (see Acknowledgements), we
reconstructed the specification of the EcoManager protocol.

With the reconstructed protocol in hand, we built our own base station
by programming an open-source, rapid-development platform called the
Nanode\cite{Nanode}.  The Nanode includes an Atmel ATmega328P
microcontroller running at 16~MHz (the same microcontroller used on
several Arduinos) and a HopeRF RFM12b\cite{RFM12b} radio frequency (RF)
module. The EcoManager products appear to use the same (or similar) RF
module tuned to the 433~MHz ISM band (note that, in some countries, it
is illegal to use the 433~MHz band without a license).  We found the
appropriate starting point for our RF configuration settings by using
a Bus Pirate\cite{BusPirate} to sniff configuration packets from the
Serial Peripheral Interface (SPI) connecting the EcoManager's
microcontroller to its RF module.

Each IAM picks its own 32-bit ID at random when the IAM's `pair'
button is pressed.  Each IAM stores its ID in non-volatile memory.
Our base station maintains a list of these IDs and polls each IAM in
order. Each IAM replies to its polling packet within 20~ms, although
the power data in this packet may be a few seconds old.  The
EcoManager RF protocol uses a modular sum checksum byte to provide
some resilience against RF corruption.  Power data is sent from our
Nanode base station to a data logging PC over an FTDI-to-USB cable.
It is also possible to turn IAMs on or off remotely.

To measure power from hard-wired appliances such as boilers and
kitchen ceiling lights, we used Current Cost transmitters\cite{CCTX}
(TX) with current transformer (CT) clamps.  These transmitters use the
same radio frequency as the EDF IAMs but a different protocol.  In
particular, the Current Cost transmitters cannot \emph{receive} RF
data.  Instead they transmit a data packet once every $6 \pm
0.3$~seconds without first checking if the RF channel is clear. Hence
RF collisions are inevitable and there is no mechanism to request
re-transmission of lost data.  As such, our base station minimises the
chance of packet collisions by learning the transmit period of each
Current Cost TX and ensuring that the base station does not transmit
for a short window of time prior to the expected arrival of a Current
Cost TX packet. Current Cost transmitters do not use a
checksum. Instead they use Manchester encoding.  RF corruption may
result in an invalid Manchester code.  If this happens then the
receiver can detect the corrupt Manchester code and discard the
packet. Unfortunately, it is possible for corruption to damage the data payload
without invalidating the Manchester encoding and so corrupt packets
are not guaranteed to be detected from Current Cost transmitters.

To maximise the distance over which we can transmit, we experimented
with several antenna and RF module configurations. We settled on a
$\sfrac{1}{4}$-wavelength antenna combined with a ground plane
composed of four $\sfrac{1}{4}$-wavelength wires in a cross shape
running in the plane of the ground, originating from just below the
point at which the antenna connects to the Nanode's printed circuit board.

\subsection{Measuring whole-house power demand}

Given that there will be a flood of smart meter data in the near
future, disaggregation researchers need access to data which is as
similar as possible to the data that will be recorded by smart meters
in the near future.  Unfortunately, `real' smart meters are not
trivial to install or to acquire: installation requires an electrician
from the utility company and, critically, the current UK smart meter
engineering specifications have not yet been finalised (the SMETS2
document\cite{SMETS2} is not an engineering specification).  British
Gas have installed over a million `real' smart meters but these meters
do not comply with SMETS2 and none of our participants had such a
meter installed.  So we had to build our own measurement system.

The current iteration of the UK smart meter specification
\cite{SMETS2} is detailed enough to allow us to build our own metering
system which closely mimics what a UK smart meter is likely to provide
(these specifications are subject to formal change control processes
such that any changes are subject to analysis and stakeholder
approval). The latest specifications\cite{SMETS2} state that smart
meters are required to connect to the Home Area Network (HAN) using
ZigBee Smart Energy Protocol v1. Presumably, a disaggregation system
would access smart meter data by way of a `Consumer Access Device'
(CAD) connected to the HAN. CADs can request instantaneous active
power and pricing data from the smart electricity meter once every
10~seconds. Additionally, CADs will be able to pull other data such as
13~months of half hourly active import data, 3~months of half hourly
reactive import data and 3~months of half hourly active and reactive
export data” [SMETS2\cite{SMETS2} and personal communication with
  DECC, October 2013].

We set out to build a metering system that would collect active
power once a second, as well as to sample the voltage and current
waveforms at 44.1~kHz to allow researchers who are interested in this
high frequency data to use it.

One solution was to use an off-the-shelf Current Cost whole-house
transmitter with a current transformer (CT) clamp. These work with our
wireless base station. We used this solution in several houses where
our bespoke solution was impractical.  There are several disadvantages
to using a CT clamp connected to a wireless transmitter:

\begin{itemize}
  \item CT clamps measure current ($I$).  The transmitter usually has
    no way to measure voltage and so must use a hard-coded value for
    voltage ($V$) to calculate a power reading ($P$) using $P=I \times
    V$. However, mains voltage in the UK is allowed to vary by
    \mbox{+10\% to -6\%} (sometimes quite abruptly) so power readings
    for a linear resistive load can vary by \mbox{+20\%
      to -12\%} (as noted by Hart\cite{Hart1992}).  These abrupt
    changes due to external noise are problematic
    for disaggregation algorithms because disaggregation algorithms
    tend to rely on \textit{changes} in power demand to detect appliance
    state changes.  However, not all appliances are affected by
    voltage variations. The power demand for `constant power'
    devices remains constant across the legal voltage range.
  \item Battery powered transmitters tend to sparsely sample from
    their CT clamp in order to minimise battery usage.  Hence rapid
    changes may be missed. 
  \item Without instantaneous measurements of both voltage and
    current, it is not possible to measure active power or reactive
    power. Hence CT clamps without voltage measurements can only
    estimate apparent power.
  \item The
    OpenEnergyMonitor
    emonTx\cite{emontx} is unaffected by all three disadvantages
    mentioned above. However, the version of the emonTx available in
    2012 when we designed our system used an analogue to digital
    converter with only 10~bits of resolution.  If we want to measure
    a primary current which varies from, say, 0 to 30~amps then the
    emonTx can only resolve changes larger than 14~watts. (The
      emonTx uses 10~bits of resolution to capture both the positive and
      negative sides of the AC signal so in effect it uses only 9~bits to
      cover a 30~amp range. $\SI{30}{\ampere} \div 2^9\,\text{ADC
        steps}=\SI{0.06}{\ampere}$ per ADC step so it can resolve
      changes in current larger than or equal to $\SI{0.06}{\ampere}$.  And
      $\SI{0.06}{\ampere} \times \SI{230}{\volt} =
      \SI{13.8}{\watt}$).  `Real' smart meters will almost certainly
    have considerably higher resolution so, unfortunately, the 
    emonTx available in 2012 when we were designing
    our system was not a suitable proxy for a `real'
    smart meter.
\end{itemize}

No existing home energy monitor that we were aware of provided an
accurate proxy for UK smart meters.  Expensive power quality monitors
costing several hundred or thousand UK pounds can measure with the accuracy we
require but these are prohibitively expensive and some require CT
sensors \emph{without} a split core, hence requiring the installer to
disconnect the meter tails from the utility company's meter, which can
only be done with permission from the utility company.

We propose a low-cost, high resolution, easy to install technique for
recording whole-house mains power demand using a computer sound
card, a CT clamp and an AC-AC adapter.

Typical sound cards have remarkably good analogue to digital
converters (ADCs).  Typical specifications of a modern sound card
include:

\begin{itemize}
  \item 96~kHz sample rate.
  \item Simultaneous recording of at least 2 channels.
  \item 90~dB signal to noise.
  \item 20~bits per sample.  Given that each bit provides 6 dB of
    dynamic range, we effectively have $\sfrac{90}{6} = 15$ bits of
    `signal' and $20-15=5$ bits of `noise' per sample.
  \item Built-in high-pass filter.
  \item Built-in anti-alias filter.
\end{itemize}

To record mains voltage and current waveforms we require a simple
circuit to connect the sound card to an AC-AC adapter and a CT clamp
(Figure~\ref{fig:system_diagram}). This circuit does not require the
user to handle any hazardous voltages.  We used the line-input of the
sound card rather than the microphone input because the line-input
should provide a lower noise signal path than the sound card's
microphone pre-amplifier.  The standard maximum peak-to-peak voltage
for consumer audio equipment line-input is 0.89~volts.  Hence the aim
of our circuit must be to reduce the output voltage of each sensor so
that we never deliver more than 0.89~volts to the sound card.

To measure mains voltage as safely as possible, we used a standard
AC-AC adapter (the `Ideal Power 77DB-06-09').  This provides a peak
open-circuit output voltage of approximately 11 volts. Research done
by the Open Energy Monitor project\cite{Wall} suggests that the output of the
AC-AC adapter should track the mains input voltage linearly over the
range 185.5~V to 253~V.
We reduced the AC-AC adapter's output voltage with a voltage divider
circuit (we used two resistors: 10~k\ohm~and 220~\ohm) to produce about
0.7 V peak-to-peak which is fed into one channel of the sound card's
line input.

To measure mains current, we used a current transformer (CT) clamp
(the `YHDC SCT-013-000'). Both the CT clamp and AC-AC adapter were
sourced from the Open Energy Monitor shop\cite{OEMshop}.  The CT is
connected in parallel to a 22~\ohm~burden resistor. This configuration
produces about 0.89~V peak-to-peak across the burden resistor when the
CT is presented with a primary current of 30 amps RMS which, we
believe, is the most current that any of the houses under study will
pull.

To protect the sound card against overload, both channels include an
80 mA quick-blow fuse and a pair of 1N5282 diodes (with a 1.3 V
forward voltage bias) to ensure that the circuit is unlikely to ever
deliver more than 1.3 V to the sound card.

Let us calculate a rough estimate for our measurement resolution.  If
we want to measure a primary current with a range of 0 to
$\SI{30}{\ampere}_{\text{rms}}$ then we should be able to resolve
changes in primary current of approximately \SI{3}{\milli\ampere} per
sample ($\SI{30}{\ampere}_{\text{rms}} \times \sqrt{2} \times
  2 \approx \SI{85}{\ampere}_{\text{peak-to-peak}}$ and
  $\SI{85}{\ampere}_{\text{peak-to-peak}} \div 2^{15}\text{ADC steps}
  \approx \SI{3}{\milli\ampere}$).  For the voltage measurement, if
we want a range of 0 to $\SI{253}{\volt}_{\text{rms}}$
($\SI{230}{\volt}_{\text{rms}} + 10\,\%$) then we should be able to
resolve changes of approximately \SI{22}{\milli\volt} per
sample ($\SI{253}{volt}_{\text{rms}} \times \sqrt{2} \times 2
  \approx \SI{716}{\volt}_{\text{peak-to-peak}}$ and
  $\SI{716}{\volt}_{\text{peak-to-peak}} \div 2^{15} \text{ADC steps}
  \approx \SI{22}{\milli\volt}$).  Given that the sensors are likely
to be noisy and given that we are only providing
$\SI{0.7}{\volt}_{\text{peak-to-peak}}$ to the ADC for the voltage
measurements, we should downgrade our resolution per sample to about
\SI{30}{\milli\volt} and \SI{5}{\milli\ampere} for voltage and
current respectively.  This gives us a resolution for power of
approximately $\SI{30}{\milli\volt} \times \SI{5}{\milli\ampere} =
\SI{150}{\milli\watt}$.

We now describe the software for our sound card power meter.  We use
the following relations to calculate $|S|$ (apparent power) and $P$
(`real' or `active' power) from simultaneously recorded vectors of
voltage and current readings (we record in chunks each with a duration
of 1~second; this time period was chosen because REDD uses this
sample period for mains data):

\begin{equation}
  |S| = I_{\text{rms}} \times V_{\text{rms}}
\end{equation}
\begin{equation}
  P = \frac{1}{N} \sum_{i=1}^{N}I_iV_{i}
\end{equation}

Where $I_{\text{rms}}$ and $V_{\text{rms}}$ are the root mean squared
values for the current and voltage vectors respectively; $N$ is the
number of samples; $I_i$ and $V_i$ are the $i^\text{th}$ samples of
the current and voltage vectors respectively.  The system does not
guarantee that we always process chunks of length equal to precise
integer multiples of the mains cycle period but, as demonstrated in
the \hyperref[sec:Validation]{Technical Validation} section, we still achieve relative errors
consistently less than 2\%.

The conclusion is that we achieve a resolution greater than that
required to provide a good proxy for `real' smart meters (although we
acknowledge that we do not know the precise resolution of `real' smart
meters.  This decision is likely to be left to the manufacturers
[personal communication with DECC, March 2013]).  We save $P$, $|S|$
and $V_{rms}$ to disk once a second with a precision of 2 decimal
places in a CSV file.

We also save the raw ADC data to disk.  To reduce the space required,
the ADC data are down-sampled using the open-source audio tool
\texttt{sox}\cite{sox} to 16~kHz (REDD used 15~kHz
and we originally wanted to use the standard defined by REDD but we
found that support for 16~kHz is more common than for 15~kHz in
processing tools).  The ADC is 20-bit but few audio processing tools
can process 20-bit files so we pad each sample to produce a 24-bit
file. The uncompressed 16~kHz 24-bit files would require 28.8~GBytes
per day so we compress the files using the Free Lossless Audio Codec
(FLAC)\cite{FLAC} to reduce the storage requirements to $\approx4.8$~GBytes per
day.

\subsubsection{Calibration}

To convert the raw ADC values to voltage and current readings, we
must first find appropriate conversion constants.  We
calibrate each data collection system separately to compensate for
manufacturing variability in the components.  We calibrate each system
once when it is first setup.  We connect a `Watts up? PRO
meter\cite{WattsUp}' to the data logging PC via USB during setup to
automatically calibrate voltage and current conversion factors. We
typically use a resistive load like a kettle to calibrate the system.
If the `Watts up?' meter reports a power factor greater than 0.97 then the
calibration script also calibrates the phase shift introduced by
the sensors.

\subsubsection{Open source implementation}

We have implemented the power monitoring system described as five
software projects. All software packages are available from \hfill \\ 
\href{https://github.com/JackKelly/}{github.com/JackKelly/}$<$package name$>$.  The 
packages are:

\begin{description}
  \item[rfm\_edf\_ecomanager] \hfill \\ Nanode C++ code. This code
    allows the Nanode to talk directly to multiple Current Cost
    whole-house sensors (CC TXs) as well as to multiple EDF
    Transmitter Plugs (CC TRXs). Users talk to the Nanode over the
    serial port. Users send simple commands. It sends data back to the
    PC in a simple JSON format.
  \item[rfm\_ecomanager\_logger] \hfill \\ A Python script for
    communicating with the rfm\_edf\_ecomanager Nanode system. This
    provides a command-line tool for `pairing' sensors with the
    logging system; assigning human-readable names to those sensors
    and then recording the data to disk in CSV files using the same
    format as MIT's REDD files. The emphasis is on reliable
    logging. rfm\_ecomanager\_logger attempts to restart the Nanode if
    the Nanode crashes. rfm\_ecomanager\_logger ensures, as far as
    possible, that time stamps are correct (which is not trivial
    given that the Nanode does not have a real time clock and given
    that serial data could be kept in the operating system's buffer if
    the system is under heavy load).  Data are recorded approximately
    once every six~seconds for each channel.
  \item[powerstats] \hfill \\ Produce statistics and graphs from
    REDD-formatted power data. Mainly used for checking the health of
    sensors.
  \item[babysitter] \hfill \\ A Python module for `babysitting' each
    logging system. Sends an email if a sensor stops working or if
    rfm\_ecomanager\_logger fails. Also sends a `heartbeat' email once
    a day to the home owner containing statistics (created by
    powerstats) describing the last day's power data.  Also provides
    useful `health' information about the system such as remaining
    disk space.
  \item[snd\_card\_power\_meter] \hfill \\ System for recording
    voltage and current waveforms at 44.1~kHz, 20-bit per channel
    using a PC's sound card. Calculates and saves active power, apparent
    power and RMS voltage to a CSV file once a second. Records
    down-sampled ADC data to a FLAC file.
\end{description}

\subsection{Complete metering setup}

To collect our own dataset, we installed the following equipment
in each house:

\begin{itemize}
  \item Multiple EDF Individual Appliance Monitors.
  \item A CurrentCost CT clamp and transmitter to measure whole-house
    apparent power. House~1 used additional CC CT clamps to measure
    the lighting circuit, kitchen ceiling lights, boiler and solar hot
    water pump.
  \item Nanode running our rfm\_edf\_ecomanager code.
  \item A small-footprint Atom PC (full component listing of the Atom
    PCs we built can be found at
    \href{http://jack-kelly.com/intel_atom_notes}{jack-kelly.com/intel\_atom\_notes}
    and a guide to setting up a complete data logging system can be
    found at
    \href{https://github.com/JackKelly/rfm_ecomanager_logger/wiki/Build-a-complete-logging-system}{github.com/JackKelly/rfm\_ecomanager\_logger/wiki/
      Build-a-complete-logging-system}). We used
    the Intel DN2800MT motherboard (with a Realtek ALC888S audio
    codec capable of sampling at 96~kHz 20-bit resolution with a
    signal to noise ratio of 90~dB; and a line-input socket on the
    rear) and a 320 GB HDD; runs Ubuntu Linux Server; consumes
    14~W active power.
  \item Houses~1, 2 and 5 had the sound card power meter system installed
    to measure whole-house active and reactive power and voltage.
\end{itemize}

A system diagram is shown in Figure~\ref{fig:system_diagram}.

CSV data files recorded by the data logging PC were transmitted to a
remote server every morning using \texttt{rsync}\cite{rsync}.  FLAC files were
transferred manually using an external hard disk every two~months.

\subsubsection{Known issues}

\begin{itemize}
  \item Each IAM draws a little power ($\text{active
    power}\approx0.9~\text{W}; \text{apparent
    power}\approx2.4~\text{VA}$).  House~1 has IAMs installed on
    almost every appliance and the correlation of the sum of
    all submeters in House~1 with the mains is 0.96.  Yet the
    proportion of energy submetered in House~1 is only 80\%.  This
    reasonably low value for the proportion of energy submetered is
    likely due in large part to the fact that the 52 EDF IAMs
    installed in House~1 draw approximately 50~W, yet this power is not
    measured by the individual appliance meters.
  \item The IAMs and the Current Cost transmitters occasionally
    report spurious readings.  rfm\_ecomanager\_logger filters out any
    readings above 4~kW for IAMs and above 20~kW for whole-house
    readings.  4~kW is above the safe maximum power draw for a UK
    mains appliance (2.99~kW = 13 amps $\times$ 230 volts).
    20~kW is more than twice the maximum whole house power reading we
    have recorded (8.765~kW) across all houses in our dataset.
  \item Whilst the Atom motherboard's ADC is capable of sampling at
    96~kHz, we had to use 44.1~kHz because the system produced buffer
    overflow errors if the sample rate was above 44.1~kHz.  There
    remain some missing samples in the 16~kHz data due to buffer
    overflow errors during recording.
\end{itemize}

\subsection{Selecting houses to record}
 
The subjects were either MSc students or PhD students at Imperial
College.  The subjects chose to do a research project with the
authors.  To assist in both their own project and in the collection of
the UK-DALE dataset, the students kindly agreed to install metering
hardware in their house.  The upper bound on the number of houses we
could record from was set by a combination of a limited financial
budget, limited time to assemble the metering hardware, and a limit in
the number of students who volunteered to work on research projects
relating to domestic energy consumption.

Within each house, the home owner selected which appliances to record,
with the recommendation from the authors that the most energy hungry
appliances should take priority.

\section{Data Records}  

UK-DALE uses a data format similar to that used by the first public
disaggregation dataset, the Reference Energy Disaggregation Data Set
(REDD)\cite{J.ZicoKolter2011}.

There are five directories, one per house.  The directories are named
\texttt{house\_<x>} where $x$ is an integer between 1 and 5.

Each directory contains a set of \texttt{channel\_<i>.dat} CSV files
(one file per electricity meter $i$) and a \texttt{labels.dat} file
which is a CSV file which maps from channel number $i$ to appliance name.
All CSV files in UK-DALE use a single space as the column separator
(as per REDD).


One way in which UK-DALE differs from REDD is that UK-DALE includes a
set of detailed metadata files. These follow the NILM Metadata
schema\cite{NILM_Metadata}.  The metadata files are in YAML text file
format (YAML is a superset of the JSON format).  This metadata
describes properties such as the specifications of each appliance; the
mains wiring between the meters and between meters and appliances;
exactly which measurements are provided by each meter; which room each
appliance belongs in etc. The \texttt{labels.dat} file in each
directory is redundant and is only included to provide compatibility
with REDD.
 
All data in UK-DALE as of January 2015 are available from the UK
Energy Research Council's Energy Data Centre.
The data are also available from
\href{http://www.doc.ic.ac.uk/~dk3810/data/}{\url{www.doc.ic.ac.uk/~dk3810/data}}.
The latter source will be updated as we collect more data.  There are
three forms of data in UK-DALE:

\begin{itemize}
  \item The 6~second data from the Current Cost meters (Data
    Citation~1: DOI:\href{http://dx.doi.org/10.5286/UKERC.EDC.000001}{10.5286/UKERC.EDC.000001}).
  \item The 1~second data from our sound card power meter (Data
    Citation~1: DOI:\href{http://dx.doi.org/10.5286/UKERC.EDC.000001}{10.5286/UKERC.EDC.000001}).
  \item The 16~kHz data recorded by our sound card power meter (Data
    Citation~2: DOI:\href{http://dx.doi.org/10.5286/UKERC.EDC.000002}{10.5286/UKERC.EDC.000002}).  The
    complete set of 16~kHz files requires 4~TBytes of storage.  The
    16~kHz data is supplied as a set of 200~MByte files. Each file
    records 1~hour of data.
\end{itemize}

The 6~second data and 1~second data are stored in CSV files, one CSV
file per meter.  The first column is a UNIX timestamp (the number of
seconds elapsed since 1970-01-01 00:00:00 UTC).  The UNIX timestamp is
UTC (Coordinated Universal Time) and hence ignores
daylight saving transitions (the UK is \texttt{UTC+0} during winter
and \texttt{UTC+1} during summer).  

The 1~second data, 6~second data and metadata are also available as a
single HDF5 binary file, ready for use with the open-source energy
disaggregation toolkit NILMTK\cite{NILMTK}.  The UK-DALE
HDF5 binary file is available from
\href{http://www.doc.ic.ac.uk/~dk3810/data/}{\mbox{\url{www.doc.ic.ac.uk/~dk3810/data}}}.

\textbf{6~second data}

For the 6~second data, the second column in each CSV file is a non-negative
integer which records power demand of the downstream
electrical load.  The file names of the 6~second data take the form of
\texttt{channel\_<X>.dat} where \texttt{X} is a positive integer (no
leading zero). There are two types of 6~second resolution meters:

\begin{enumerate}
  \item Individual appliance monitor
transmitter plugs that record \emph{active} power (in units of watts).
  \item Current
transformer meters that record \emph{apparent} power (in units of volt-amperes).
\end{enumerate}

Individual appliance monitors have a push-button switch to allow users
to turn the connected appliance on and off.  We record the activity of
this switch in a \texttt{channel\_<X>\_button\_press.dat} file.  If
the switch has just been toggled \textit{on} then a `1' is recorded.  If the
switch has just been toggled \textit{off} then a `0' is recorded. The
motivation behind logging switch press events is that these provide
(imperfect) room occupancy information.

Switch \emph{on} events should be a perfectly clean recording
(i.e. the only possible reason for an on-switch event appearing in the
data is that the user pressed the switch). Unfortunately, off-switch
events may include false positives. Occasionally IAMs turn off
spontaneously (an event which is impossible to distinguish from a
genuine button press). Also, if power is lost and returned to the IAM
within 12~seconds then this will be logged as an off-switch event. If
the power is off for more than 12~seconds then the system assumes that
the IAM was deliberately unplugged and hence the system will switch
the IAM to its previous power state when it reappears; this automatic
switch event is not recorded.\\

\textbf{1~second data}

There are four columns in each CSV file recording the
whole-house power demand every second:

\begin{enumerate}
  \item UNIX timestamp.
  \item Active power (watts).
  \item Apparent power (volt-amperes).
  \item Mains RMS voltage.
\end{enumerate}

All four columns record real numbers (not integers).  The first column
has one decimal place of precision; the other columns have two decimal
places of precision.  The 1~second data is in a CSV file called
\texttt{mains.dat} in directories \texttt{house\_1},
\texttt{house\_2} and \texttt{house\_5}.\\

\textbf{16~kHz data}

The 16~kHz data is compressed using the Free Lossless Audio Codec
(FLAC)\cite{FLAC}.  For houses 1, 2, and 5 UK-DALE records a stereo
16~kHz audio file of the whole-house current and voltage
waveforms. The files are labelled \texttt{vi-<T>.flac} where $T$ is a
real number recording the UNIX timestamp with micro-second precision
(using an underscore as the decimal place).  This timestamp is the
time at which the audio file began recording.  The recordings are
split into hour-sized chunks.  We also include a \texttt{calibration.dat}
file for each house. This is a text file specifying the multipliers
required to convert the raw output of the analogue to digital
converter to amps and volts.

To make use of the FLAC files (for processing in, for example, MATLAB
or Python), first decompress the files to create WAV files. This
decompression can be done with many audio tools.  We use the audio
tool \texttt{sox}\cite{sox}.

With the WAV files in hand, the next task is to convert from the
values in the WAV files (in the range $[-1,1]$) to volts and amps. Use
the \texttt{calibration.cfg} file for the house in question. This file
specifies an \texttt{amps\_per\_adc\_step} parameter and a
\texttt{volts\_per\_adc\_step} parameter.  Users can safely ignore the
\texttt{phase\_difference} parameter and assume that the measurement
hardware introduces no significant phase shift. Use the following
formula to calculate volts from the WAV files:
$$ \text{volts} = \text{value from WAV} \times \text{volts per ADC step} \times
2^{31} \text{ADC steps}$$

Use a similar formula for amps. To explain the formula above: The
recording software stores each sample as a 32~bit integer. Hence there
are $2^{32}$~ADC steps for the full range from $[-1,1]$ and
$2^{31}$~ADC steps for half the range. 

\label{sec:Validation}
\section{Technical Validation}

\begin{table*}
  \begin{tabular}{|p{6cm}|p{2cm}|p{2cm}|p{2cm}|p{2cm}|p{2cm}|}\hline%
    \bfseries House & \bfseries 1 & \bfseries 2 & \bfseries 3 &
    \bfseries 4 & \bfseries 5
    \begin{filecontents*}{table_1.csv}
House,1,2,3,4,5
Building type,end of terrace,end of terrace,,mid-terrace,flat
Year of construction,1905,1900,,1935,2009
Energy improvements,solar thermal \& loft insulation \& solid wall insulation \& double glazing,cavity wall insulation \& double glazing,,loft insulation \& double glazing,
Heating,natural gas,natural gas,,natural gas,natural gas
Ownership,bought,bought,,bought,bought
Number of occupants,4,2,,2,2
Description of occupants,2 adults and 1 dog started living in the house in 2006. One child born in 2011. Second child born in 2014.,2 adults. 1 at work all day; the other sometimes home,,1 adult and 1 pensioner,2 adults
Total number of meters,54,20,5,6,26
Number of site (mains) meters,2,2,1,1,2
Sample rate of mains meters,16 kHz \& 1 Hz \& 6 seconds,16 kHz \& 1 Hz \& 6 seconds,6 seconds,6 seconds,16 kHz \& 1 Hz \& 6 seconds
Date of first measurement,2012-11-09,2013-02-17,2013-02-27,2013-03-09,2014-06-29
Date finished installing all meters,2013-04-12,2013-05-22,“,“,“
Date of last measurement,2015-01-05,2013-10-10,2013-04-08,2013-10-01,2014-11-13
Date when some meters were removed,,,,,2014-09-06
Total duration (days),786,234,39,205,137
Total uptime for mains meter (days),655,140,36,155,131
Uptime proportion,0.83,0.60,0.93,0.75,0.96
Average mains energy consumption per day (active kWh),7.64,7.17,,,13.75
Average mains energy consumption per day (apparent kVAh),8.90,8.00,12.35,10.24,17.56
Following statistics calculated when all meters installed,,,,,
Correlation of sum of submeters with mains,0.96,0.86,0.47,0.55,0.90
Proportion of energy submetered,0.80,0.68,0.19,0.28,0.79
Mean dropout rate (ignoring large gaps),0.02,0.02,0.02,0.02,0.02
\end{filecontents*}
\csvreader[head to column names]{table_1.csv}{}%
    {\\\hline\House & \1 & \2 & \3 & \4 & \5}
    \\\hline
  \end{tabular} 
  \caption{\textbf{Summary statistics for each house.} `uptime' is the
    total time that the system was active and recording.  The `total
    duration' is `date of last measurement' minus `date of first
    measurement'. The correlation of the mains meter with the sum of
    all submeters gives an indication of how much of the variance in
    the mains signal is captured by the submeters.  The proportion of
    energy submetered is the total energy captured by the submeters
    divided by the total energy captured by the mains meter.  The
    dropout rate (ignoring large gaps) gives a measure of the rate at
    which packets were lost due to radio errors (large gaps are
    ignored because these are often caused by a meter being
    deliberately unplugged). Some metadata is not available because
    the occupants are no longer contactable.}
  \label{table:houses}
\end{table*}

Table~\ref{table:houses} summarises the UK-DALE dataset.  The table
includes some metadata (which is also recorded in the machine-readable
metadata supplied with the dataset) including the type of building,
the year of construction, the main heat source, whether the property
is bought or rented, the number of occupants, a description of the
occupants, the total number of meters, the number of site meters, the
sample rate of the mains meters and the start and end dates for the
recordings. The table also includes summary statistics calculated
using the open source energy disaggregation tool NILMTK\cite{NILMTK}:
the average mains energy consumed per day, the correlation of the
mains meter with the sum of all submeters, the proportion of energy
submetered, and the dropout rate.  The values for the average energy
consumption per day are close to the value of 9.97~kWh per day
reported in DECC's Household Electricity Survey\cite{HES} (which
surveyed 251 houses in the UK), hence we can have some confidence that
our houses consumed a fairly typical amount of energy for a UK house.

Figures~2 to 6 (except panel `a' in Figure~4) were produced using
NILMTK.  The scripts to generate these plots are available at
\href{https://github.com/JackKelly/ukdale_plots}{github.com/JackKelly/ukdale\_plots}

\begin{figure*}
  \centering
  \ifscientificdata
    02\_area\_plot.eps
  \else
    \includegraphics{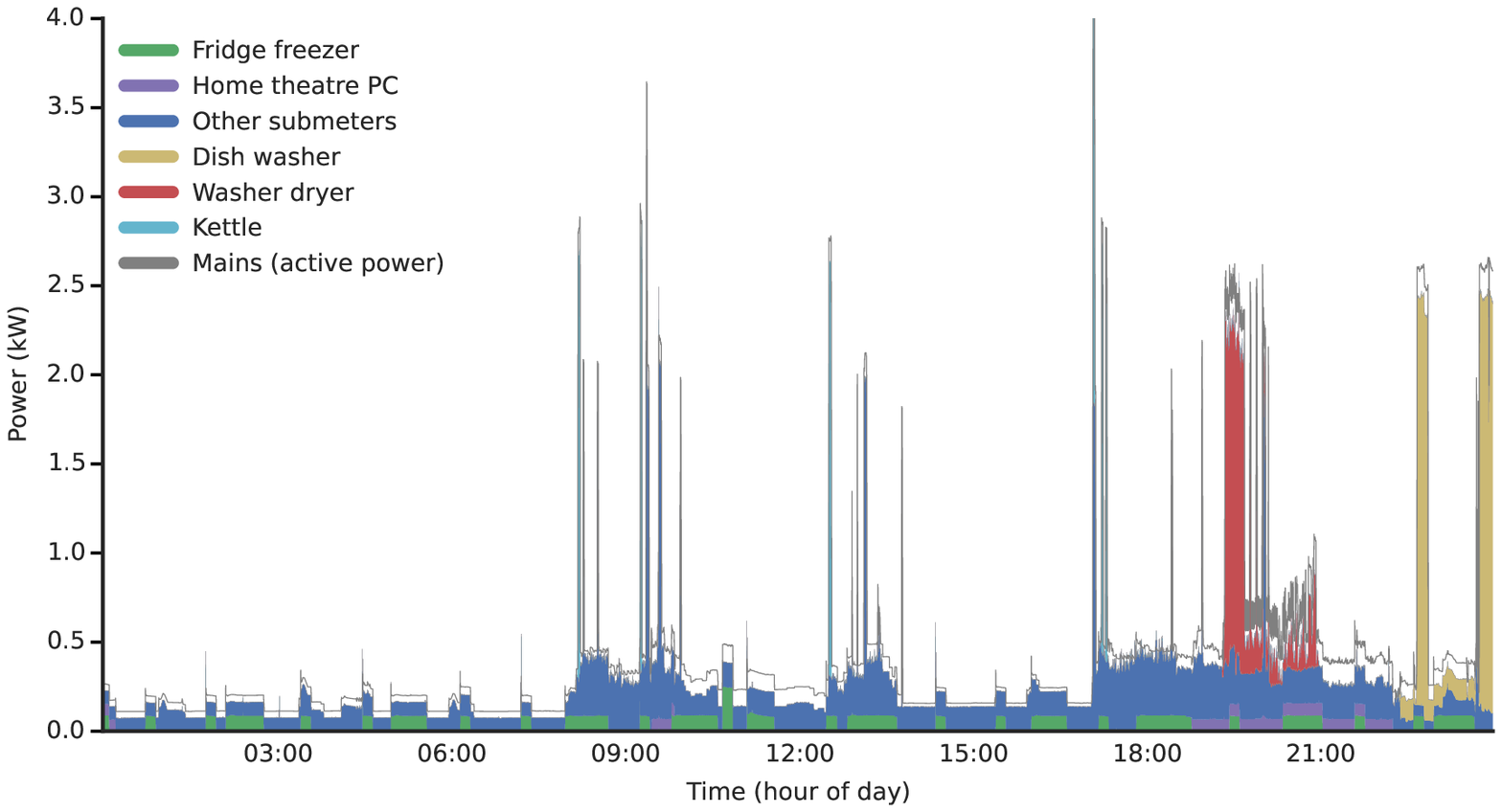}
  \fi
  \caption{\textbf{Power demand for a typical day (Sunday 2014-12-07)
      in House~1.}  The thin grey line shows the mains (whole-house)
    active power demand recorded using our sound card power meter.
    The stacked and filled coloured blocks show the power demand for
    the top five appliances (by energy consumption) and the dark blue
    block shows all the other submeters summed together.  The thin
    white gap between the top of the coloured blocks and the mains
    plot line represents the power demand not captured by any
    submeter.}
  \label{fig:area_plot} 
\end{figure*} 

Figure~\ref{fig:area_plot} shows the power demand for a typical day
for House~1.  We show the individual power demand for the top-five
appliances (ranked by energy consumption) and all other submeters
summed together.  We also show the whole-house mains power demand.
The difference between the mains power demand and the top of the
submetered power demand illustrates the small amount of energy which
is not submetered.

\begin{figure*}
  \centering
  \ifscientificdata
    03\_good\_sections.eps
  \else
    \includegraphics{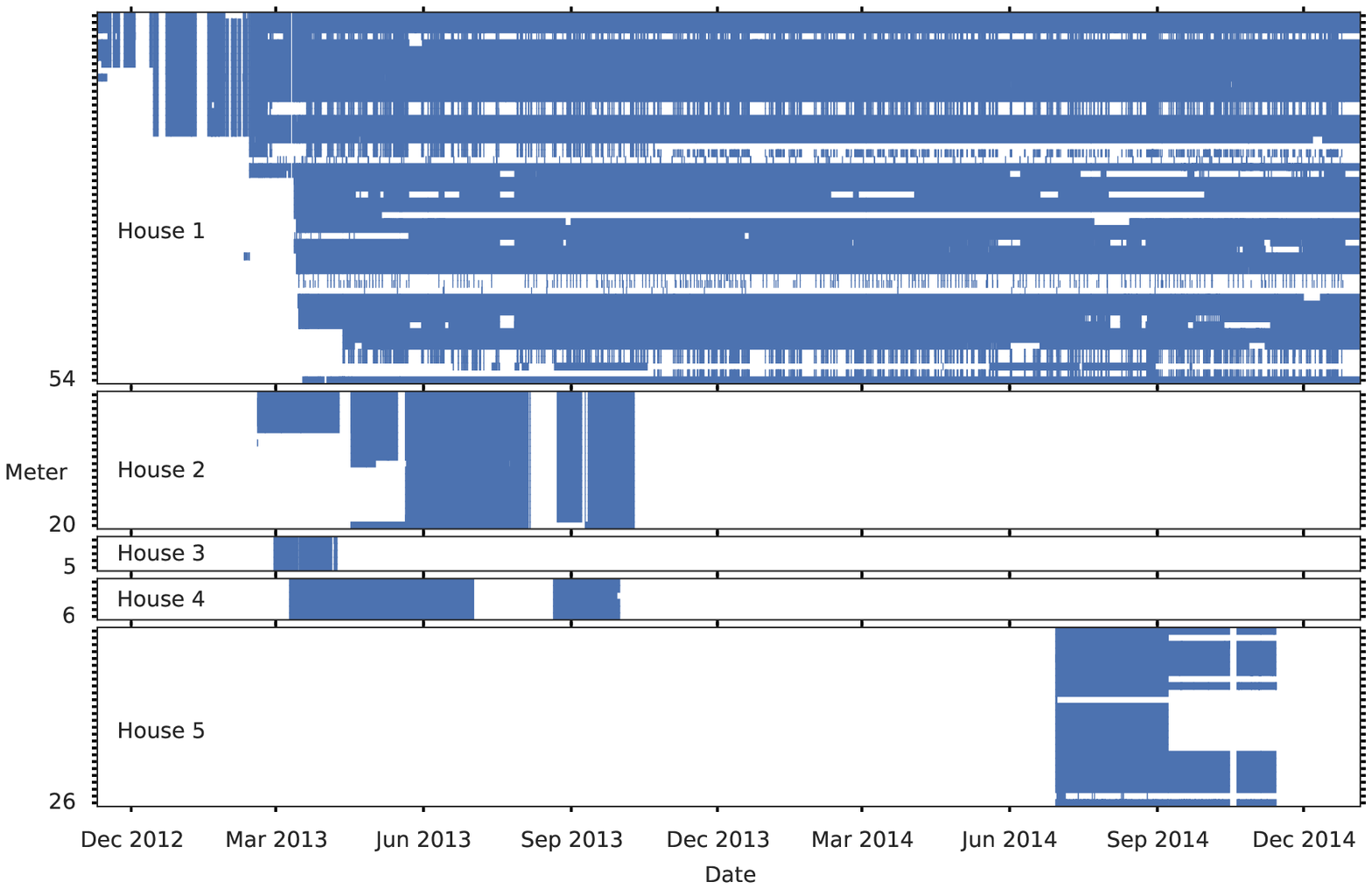}
  \fi
  \caption{\textbf{Time periods when meters were recording.}  The five
    houses in the dataset are represented by the five panels in this
    plot.  The height of each panel is proportional to the number of
    meters installed in each house.  Each thin row (marked by each
    y-axis tick mark) represents a meter.  Blue areas indicate time
    periods when a meter was recording.  White gaps indicate gaps in
    the dataset.}
  \label{fig:good_sections} 
\end{figure*}

The time periods when each meter was capturing data is shown in
Figure~\ref{fig:good_sections}. Note that the numerous gaps in the
data from House~1 are almost all deliberate and not the result of an
equipment failure.  For example, some meters in House~1 are manually
turned off if the attached appliance is unplugged.

\begin{figure*}
  \centering
  \ifscientificdata
    04\_16kHz\_and\_mains\_histograms.eps
  \else
    \includegraphics{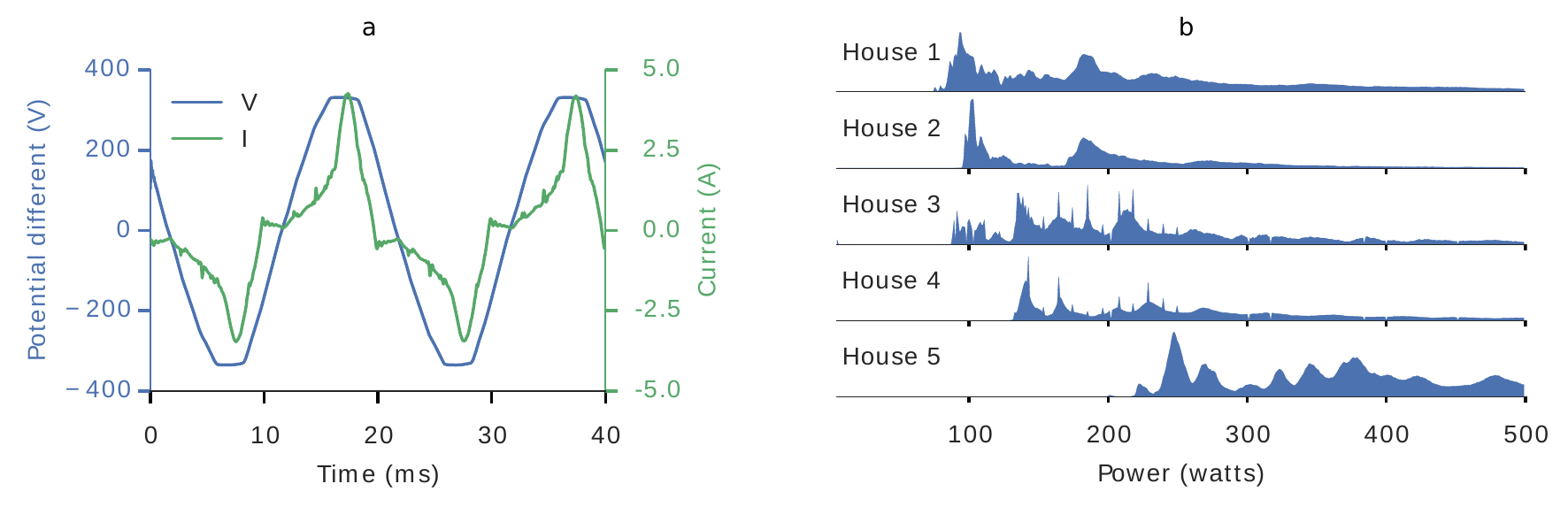}
  \fi
  \caption{\textbf{Mains electricity data.} \textbf{a,} 16~kHz
    sampling of mains voltage and current using our sound card power
    meter from House~1 on 2014-09-03 21:00:00+01:00.  The green
  line shows the current and the blue line shows the voltage.  Panel
  \textbf{b} shows histograms of mains power demand for each house. 
    The five subplots represent the five houses in the dataset.  There
    is some density above 500~watts but this has been cropped from
    this plot to allow us to see detail in the range between 0 and
    500~watts.}
  \label{fig:16kHz}
\end{figure*}

An example of 16~kHz data captured by our sound card power meter is
shown in Figure~\ref{fig:16kHz} panel `a'.  Note that the voltage is
almost a pure 50~Hz sine wave but the current contains many harmonics.

The distribution of values for the mains power demand for each house
is shown in Figure~\ref{fig:16kHz} panel `b'.  The left-most
edge of each density represents the `vampire power' of each house
(i.e. the power demand when no one is using an appliance but power is
still being drawn by always-on appliances and appliances in standby
mode).

\begin{figure*}
  \centering
  \ifscientificdata
    05\_top\_5\_energy\_and\_appliance\_activity\_histograms.eps
  \else
    \includegraphics{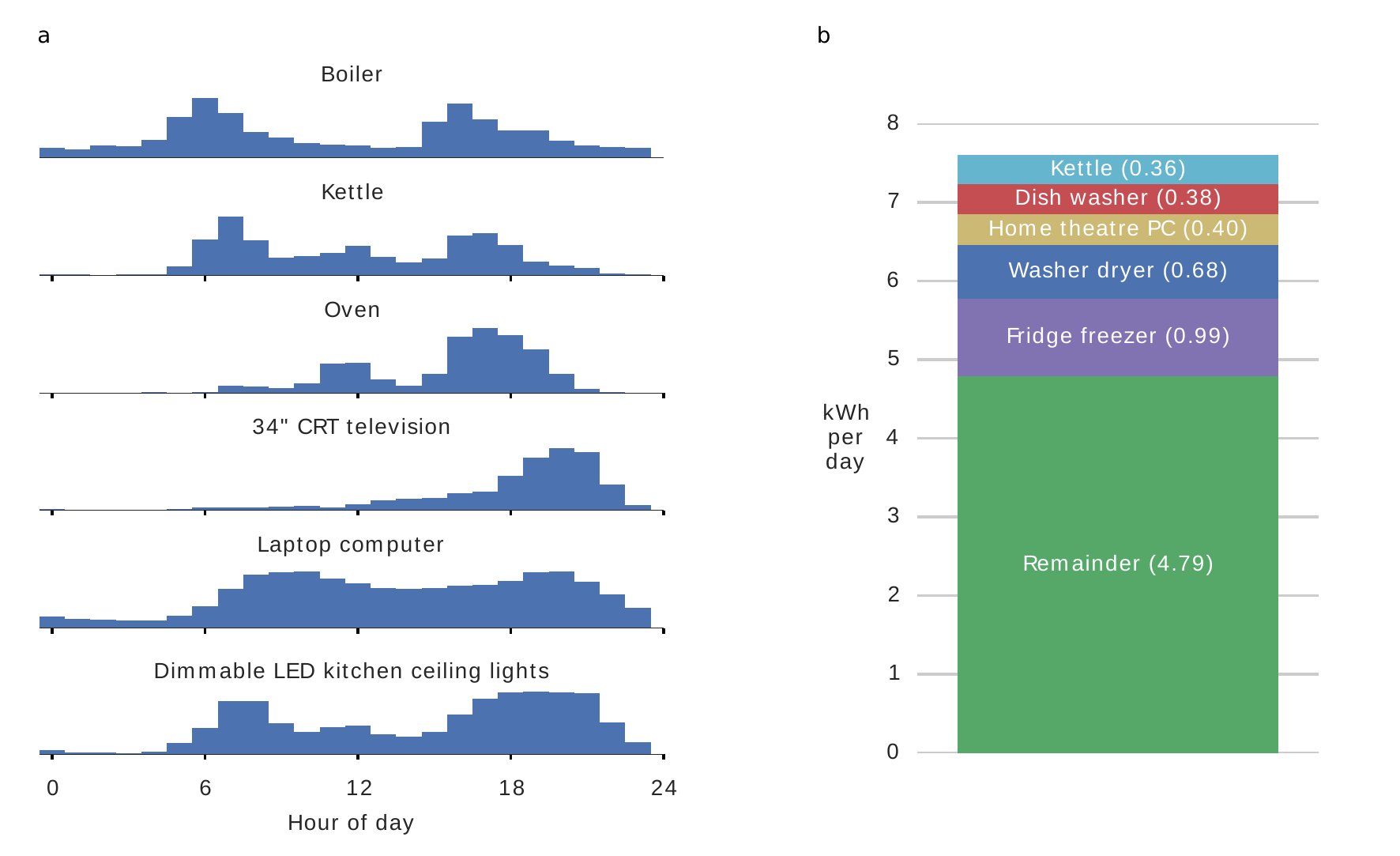}
  \fi
  \caption{\textbf{Electrical appliance usage in House~1. a,}
    Histograms of daily appliance usage patterns. Panel \textbf{b}
    shows average daily energy consumption of the top-five appliances
    in House~1.  All appliances were ranked by the amount of energy
    they consumed and the top-five are shown here.  All lights were
    grouped together.  The `remainder' block at the bottom represents
    the difference between the total mains energy consumption and the
    sum of the energy consumption of the top five appliances.  As
    such, the top edge of the bar shows the average daily total energy
    consumption for House~1.}
  \label{fig:top_5_energy} 
\end{figure*}

Figure~\ref{fig:top_5_energy} panel `a' shows the hour per day
that several appliances are used.  For example, the oven shows two
peaks in usage: one around midday (lunch) and one around 18:00
(dinner).

The energy consumed by the top five energy consuming appliances in
House~1 is shown in Figure~\ref{fig:top_5_energy} panel `b'.  This is
relevant because energy disaggregation researchers often prioritise
the disaggregation of the appliances responsible for the largest
energy consumption.

\begin{figure*}
  \centering
  \ifscientificdata
    06\_appliance\_power\_histograms.eps
  \else
    \includegraphics{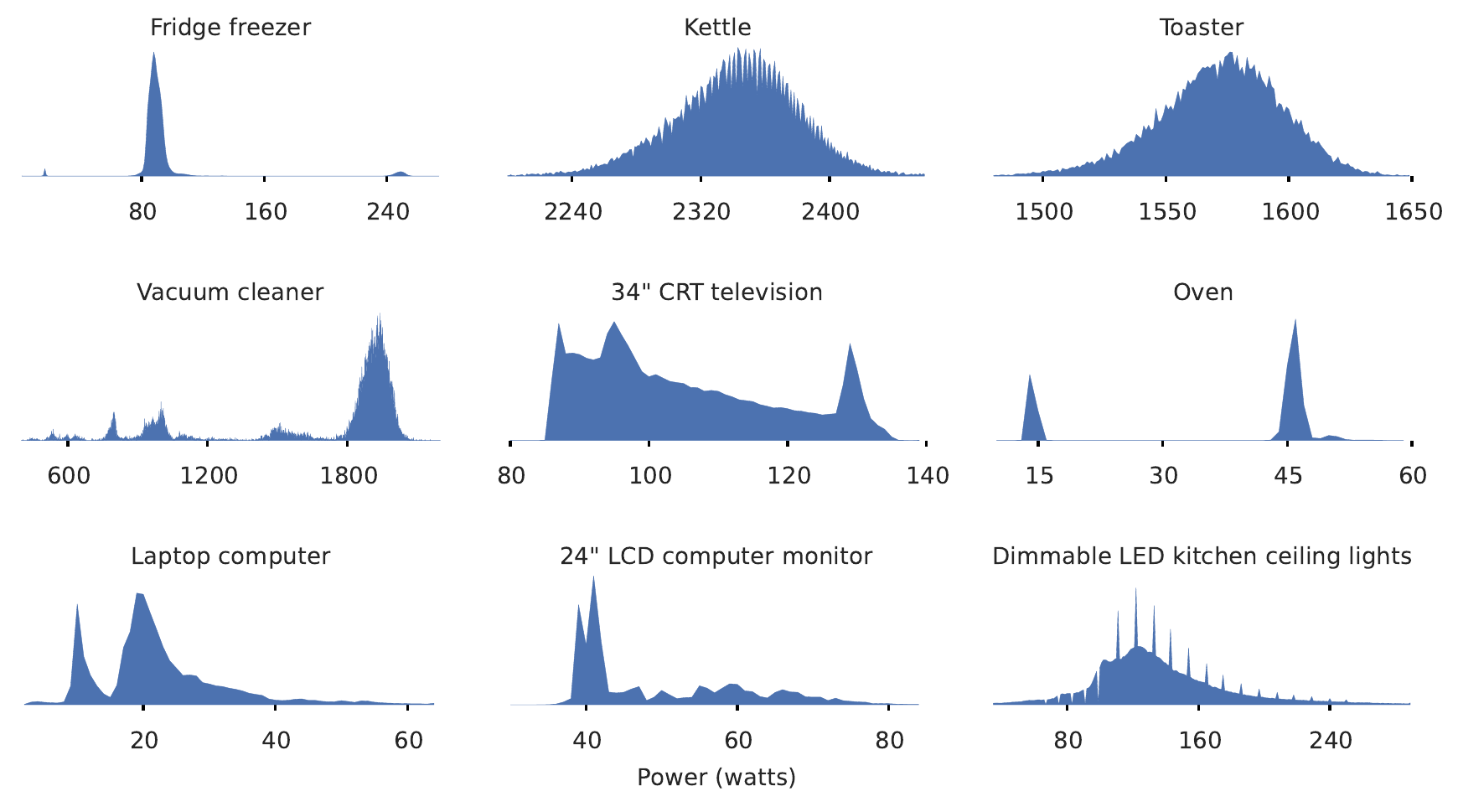}
  \fi
  \caption{\textbf{Histograms of appliance power demand from House~1.}}
  \label{fig:appliance_power_histograms} 
\end{figure*}

The distribution of values of power demand for individual appliances
is shown in Figure~\ref{fig:appliance_power_histograms}.  Some
appliance information can be inferred from the histograms.  For
example, the top left panel shows a histogram for the power demand of
the fridge: the main peak around 90~W is the normal compressor cycle,
the peak around 17~W is the fridge lamp and the peak around 250~W is
the defrosting cycle.  The vacuum cleaner has six discrete power
settings, all of which can be seen in its histogram. 

\begin{figure*}
  \centering
  \ifscientificdata
    07\_measurement\_error.eps
  \else
    \includegraphics{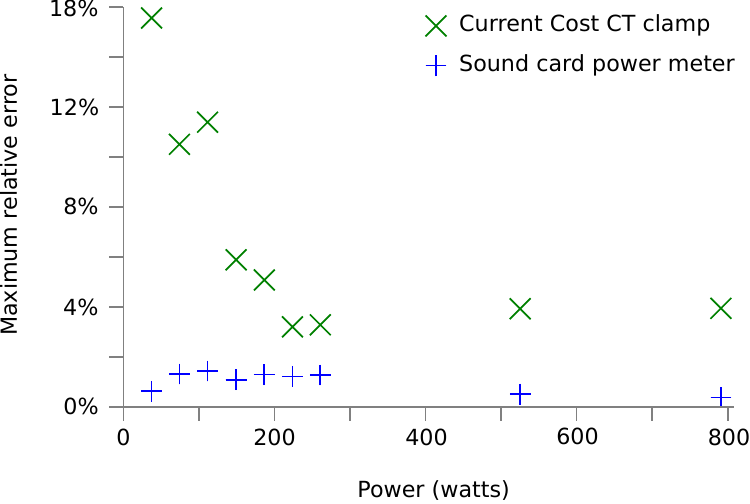}
  \fi
  \caption{\textbf{Maximum relative measurement error for power
      measurements across a range of loads.}  A `Watts up? PRO' meter\cite{WattsUp} was
    used to record the ground-truth.}
  \label{fig:measurement_error}
\end{figure*}

Figure~\ref{fig:measurement_error} shows the measurement errors for
our sound card power meter and the Current Cost Current Transformer
(CT) sensor across a range of resistive loads. The ground truth was
measured using a `Watts up? PRO' meter\cite{WattsUp} (with a stated accuracy of
$\pm1.5\%$). The range of loads were created by using three incandescent
lamps and by changing the number of primary turns on the CT from one
to seven, in steps of one.  For each load, we recorded one minute of
data and took the largest error from that minute of data.  The sound
card power meter was calibrated six months prior to the test.  This
illustrates that our sound card power meter consistently produces a
relative error of less than 2\% and that the Current Cost CT meter
produces errors of less than 6\% as long as the power is above 100
watts (the whole-house power demand very rarely drops below 100
watts).

We compared the total active energy recorded by our sound card power
meter for House~1 with the utility-installed `spinning disk'
electricity meter in House~1.  We selected the time period of
2014-11-28 08:55 to 2013-05-22 20:36 for the comparison
because we have a continuous recording from our sound card power meter
for this period, and we have readings of the utility meter at the
start and end of this period.  The total energy recorded by the
utility meter for this period was 4030.60~kWh.  The total recorded by
the sound card power meter was 4142.93~kWh. The relative difference is
2.71\%.

Houses 1, 2 and 5 had two mains meters: a sound card power meter and a
Current Cost meter. For each of these houses, we compared the apparent
energy recorded by the sound card power meter against the Current Cost
whole-house meter.  The relative difference was 1.79\% in House~1;
7.30\% in house 2 and 5.68\% in house 5.

\section{Usage Notes}

Any software designed to process REDD files should be able to open
UK-DALE data (although the extra metadata provided by UK-DALE will be
ignored).

The open-source energy disaggregation toolkit NILMTK\cite{NILMTK}
includes an importer for UK-DALE data.  NILMTK can handle the metadata
provided with UK-DALE.  An HDF5 version of the UK-DALE dataset (for
use with NILMTK) is available for download (please see the Data
Records section of this paper).

There are several aspects of the dataset that might need to be
addressed using appropriate pre-processing:

Data packets from the wireless meters are occasionally lost in
transmission.  Around 6\% of packets are lost from the Current
Transformer sensors and around 0.02\% of packets are lost from
Individual Appliance Monitor plugs.  The sample period for the 6
second data may drift up or down by a second.

Some individual appliance monitors are switched off (and hence do not
transmit any data) when the appliance is switched off from the mains.
We switch them off to reduce the risk of an electrical fault causing a
fire, and to save energy.  Some appliances are unplugged for the
majority of the time.  For example, the vacuum cleaner is physically
unplugged from the wall socket when not in use and the vacuum
cleaner's meter is left attached to the cleaner's power plug (and
hence is not powered).  As a rule of thumb, any gap in the data longer
than two minutes can be assumed to be caused by the appliance (and
monitor) being switched off from the mains.  Hence gaps longer than
two minutes can safely be filled with zeros.  The threshold of two
minutes was chosen because we observed gaps less than two minutes
caused by a succession of radio transmission errors.  Any gap shorter
than two minutes can be forward-filled from the previous reading.

Some appliances draw more than $0$~watts when
nominally `off'.  We typically use 5~watts as the threshold between
`on' and `off'.  The metadata includes an
\texttt{on\_power\_threshold} property for each appliance.  This property is
present if the on power threshold for that appliance is not 5~watts.

NILMTK contains preprocessing tools for handling these scenarios.

\section{Acknowledgments}
\begin{itemize}
  \item Geoff Dutton at the STFC (and data manager for the UKERC EDC).
  \item Dr Mark Bilton at Imperial's Low Carbon London Lab.
  \item Robert Wall and the Open Energy Monitor project.
  \item Peter Morgan from DECC.
  \item Graham Murphy, Matt Thorpe and Paul Cooper 
    who contributed hugely to the effort to
    decode the Current Cost and EDF RF protocols.
  \item The Current Cost company.
  \item This work was funded by the EPSRC and by Intel via their
  EU Doctoral Student Fellowship Programme.
  \item We thank the students whose homes we recorded from.
  \item Jack Kelly had full access to all the data in the study and
    takes responsibility for the integrity of the data and the
    accuracy of the data analysis.
\end{itemize}

\section{Author Contributions}
Jack Kelly built the hardware, wrote the software and wrote the
majority of the paper.

William Knottenbelt provided conceptual guidance as supervisor of Jack
Kelly's PhD, helped recruit MSc students for UK-DALE and provided
editorial feedback on the paper.

\section{Competing financial interests}
The authors declare no competing financial interests.



\section{Data Citations}

\begin{enumerate}[1.]
  \item Kelly, J. \& Knottenbelt, W. \textit{UKERC Energy Data Centre}
    DOI:\href{http://dx.doi.org/10.5286/UKERC.EDC.000001}{10.5286/UKERC.EDC.000001}
    (2015).
  \item Kelly, J. \& Knottenbelt, W. \textit{UKERC Energy Data Centre}
    DOI:\href{http://dx.doi.org/10.5286/UKERC.EDC.000002}{10.5286/UKERC.EDC.000002}
    (2015).
\end{enumerate}

\end{multicols}
\end{document}